\documentclass[prc,a4paper,twocolumn,amsmath,amssymb,showpacs,showkeys,floatfix]{revtex4}
\usepackage[latin1]{inputenc}
\usepackage[T1]{fontenc}
\usepackage{units}
\usepackage{graphicx}
\usepackage{dcolumn}

\begin{document}

\title{Impact of the quenching of $g_{\rm A}$ on the sensitivity of $0\nu\beta\beta$ experiments}
\author{Jouni Suhonen}
\affiliation{University of Jyvaskyla, Department of Physics,  
P.O.\ Box 35, FI-40014 Jyvaskyla, Finland}

\begin{abstract}
Detection of the neutrinoless $\beta\beta$ ($0\nu\beta\beta$) decay is of high priority
in the particle- and neutrino-physics communities. The detectability of this
decay mode is strongly influenced by the value of the weak axial-vector coupling 
constant $g_{\rm A}$. The recent nuclear-model analyses of $\beta$ and $\beta\beta$ decays 
suggest that the value of $g_{\rm A}$ could be dramatically quenched, reaching ratios of
$g^{\rm free}_{\rm A}/g_{\rm A}\approx 4$, where $g^{\rm free}_{\rm A}=1.27$ is the free,
neutron-decay, value of $g_{\rm A}$. The effects of this quenching appear devastating 
for the sensitivity of the present and future $0\nu\beta\beta$ experiments since 
the 4$th$ power of this ratio scales the $0\nu\beta\beta$ half-lives. 
This, in turn, could lead to some two orders of magnitude less sensitivity for the 
$0\nu\beta\beta$ experiments. In the present article it is shown that by using a 
consistent approach to both the two-neutrino $\beta\beta$ and $0\nu\beta\beta$ 
decays by the proton-neutron quasiparticle random-phase approximation (pnQRPA), 
the feared two-orders-of-magnitude reduction in the sensitivity of the 
$0\nu\beta\beta$ experiments actually shrinks to a reduction by factors in 
the range $2-6$. This certainly has dramatic consequences for the potential to
detect the $0\nu\beta\beta$ decay.
\end{abstract}

\pacs{21.60.Jz, 23.40.Hc, 27.50.+e, 27.60.+j}
\keywords{Detectability of the neutrinoless double beta decay, nuclear matrix elements,
quenching of the axial-vector coupling strength, sensitivity of the double-beta experiments}

\maketitle

\section{Introduction \label{sec:intro}}

The nuclear matrix elements (NMEs) of neutrinoless $\beta\beta$ ($0\nu\beta\beta$) decay
are becoming an increasingly hot issue in present-day neutrino and particle physics
since they are the ones that relate the absolute mass scale of the neutrino (the 
\emph{effective} neutrino mass $\langle m_{\nu}\rangle$) to the constantly increasing sensitivity 
of the present and future $0\nu\beta\beta$ experiments \cite{REPORT,Vergados2012,Vergados2016}.
The formidable progress in the sensitivity of the neutrino-mass related underground
experiments is witnessed in the rapidly increasing accuracy in the measured half-lives
of the two-neutrino $\beta\beta$ ($2\nu\beta\beta$) decays for some ten nuclear 
systems \cite{Barabash2010,Barabash2015}. At the same time the lower limits for
the half-lives of the $0\nu\beta\beta$ decay mode keep steadily increasing 
\cite{Vergados2016}. Important nuclei for the present and
future $0\nu\beta\beta$ experiments are
$^{76}$Ge (the GERDA experiment \cite{Agostini2017}, and in the future GERDA and 
Majorana \cite{Cuesta2015}),
$^{82}$Se (the NEMO-3 experiment \cite{Arnold2004}, and in the future
SuperNEMO \cite{Hodak2015} and MOON \cite{Fushimi2010}),
$^{96}$Zr (the NEMO-3 experiment \cite{Argyriades2010}),
$^{100}$Mo (the NEMO-3 experiment \cite{Arnold2015}, and in the future
AMoRE \cite{Park2015}, LUMINEU \cite{Becker2016}, 
CUPID \cite{Artusa2017}, MOON \cite{Fushimi2010}),
$^{116}$Cd (the NEMO-3 experiment \cite{Arnold2017}, and in the future
AURORA \cite{Danevich2016}, COBRA \cite{Zatschler2015}),
$^{130}$Te (the CUORE experiment \cite{Alfonso2015}, and in the future CUORE and 
SNO+ \cite{Andringa2016}), and
$^{136}$Xe (the EXO \cite{Albert2014} and KamLAND-Zen \cite{Gando2016} experiments, and in
the future NEXT \cite{Martin-Albo2016} and PandaX-III \cite{Chen2017}).

As mentioned above, the exact sensitivity of the present and future $0\nu\beta\beta$
experiments to the effective neutrino mass $\langle m_{\nu}\rangle$ can be determined
only if the involved NMEs are known accurately enough. For this reason, more and
more attention is being paid to the issue of computing the values of these NMEs
in a proper way \cite{Engel2017}. 

The article is organized as follows: In 
Sec.~\ref{sec:theory} the theory background of the $0\nu\beta\beta$ is briefly
outlined and in Sec.~\ref{sec:background} the background of the $g_{\rm A}$ problem is
discussed. In Sec.~\ref{sec:model} the adopted nuclear model and its properties
are introduced, as also the fitting procedure for consistent prediction of the
$0\nu\beta\beta$ NMEs. In Sec.~\ref{sec:results} the results for the computed
half-lives are given and in Sec.~\ref{sec:conclusions} the final conclusions are
drawn.

\section{Theoretical background \label{sec:theory}}

By assuming the neutrino-mass mechanism to be the dominant one, one can 
write the connection between the $0\nu\beta\beta$ half-life and the involved NMEs
conveniently as
\begin{equation} \label{eq:half}
t_{1/2}^{(0\nu)} = F^{(0\nu)}(g_{\rm A})\Bigl(\langle m_{\nu}\rangle [\textrm{eV}]\Bigr)^{-2} \, ,
\end{equation}
where $\langle m_{\nu}\rangle$ is the effective neutrino mass \cite{REPORT} in units of
eV and $F^{(0\nu)}$ can be coined \emph{reduced} half-life:
\begin{equation} \label{eq:F}
F^{(0\nu)}(g_{\rm A}) = \left( g_{\rm A}^2\left\vert M^{(0\nu)}\right\vert\right)^{-2}
\left(G_0^{(0\nu)}\right)^{-1}\Bigl( m_{e}[\textrm{eV}]\Bigr)^2 \, .
\end{equation}
Here $m_{e}$ is the electron rest mass in units of eV, $G_{0}^{(0\nu)}$
is the leptonic phase-space factor defined in \cite{Kotila2012}, and $g_{\rm A}$ is
the weak axial-vector coupling constant with the free value $g^{\rm free}_{\rm A}=1.27$,
determined from the decay of a neutron to a proton \cite{Markisch2014}. This value
is to some extent protected in nuclear medium by the PCAC (partially conserved axial-vector
current) hypothesis. The $0\nu\beta\beta$ nuclear matrix element, $M^{(0\nu)}$,
consists of the Gamow--Teller (GT), Fermi (F) and tensor (T) parts as
\begin{equation} \label{eq:NME}
M^{(0\nu)} = M_{\rm GT}^{(0\nu)} - \left( \frac{g_{\rm V}}{g_{\rm A}}
\right)^{2} M_{\rm F}^{(0\nu)} + M_{\rm T}^{(0\nu)} \ .
\end{equation}

\section{Background of the $g_{\rm A}$ problem \label{sec:background}}

Whereas the NME $M^{(0\nu)}$ of (\ref{eq:NME}) has been under intensive theoretical
discussion for decades, the other involved quantity, $g_{\rm A}$, has remained an
innocent by-stander, not provoking heated discussions in the $0\nu\beta\beta$
community. Attempts to deal with the $g_{\rm A}$-quenching problem include the shell-model
calculations \cite{Brown1978,Chou1993,Brown1988} in the 0p and 1s-0d shells and
\cite{Martinez1996,Siiskonen2001,Kumar2016} in the 1p-0f shell and beyond. The review
\cite{Towner1997} gives a comprehensive account of the $g_{\rm A}$ problem.

Only recently the possibly decisive role of $g_{\rm A}$ in the half-life
(\ref{eq:half}) and in the discovery potential of the $0\nu\beta\beta$ experiments
has surfaced \cite{Barea2013}. In \cite{Barea2013} a comparison of the experimental
and computed $2\nu\beta\beta$ half-lives of a number of nuclei yielded the striking 
result
\begin{equation}
\label{eq:curves}
 g_{\rm A}^{\rm eff}(\textrm{IBM-2}) = 1.269A^{-0.18} \ ; \quad 
g_{\rm A}^{\rm eff}(\textrm{ISM}) = 1.269A^{-0.12} \,,
\end{equation}
where $A$ is the mass number and IBM-2 stands for the microscopic interacting boson
model and ISM  is an acronym for the interacting shell model. The result (\ref{eq:curves})
implies that strongly quenched effective values of $g_{\rm A}$, 
in the range $g^{\rm eff}_{\rm A}=0.5-0.7$ for 
$A=76-136$, are possible, thus affecting substantially the discovery potential of the
$0\nu\beta\beta$ experiments.

Although the study \cite{Barea2013} was the first to draw considerable
attention in the experimental
$0\nu\beta\beta$ community, it was not the first one to point to a possible strong
quenching of $g_{\rm A}$. Already the study \cite{Faessler2008} gave indications of a
heavily quenched effective $g_{\rm A}$, in the range 
$g^{\rm eff}_{\rm A}=0.39-0.84$. In \cite{Faessler2008}
the Gamow-Teller $\beta$ decays and $2\nu\beta\beta$ decays in the $A=100,116,128$
$\beta\beta$-decay triplets  were studied by the use of the proton-neutron
quasiparticle random-phase approximation (pnQRPA), and a least-squares procedure
was used to optimize the values of the experimentally over-constrained parameters
$g_{\rm A}$ and $g_{\rm pp}$, where $g_{\rm pp}$ is the interaction strength of the
particle-particle part of the proton-neutron interaction \cite{Vogel1986,Civitarese1987}.
Also in the ISM studies \cite{Juodagalvis2005,Caurier2012} very low values of 
$g_{\rm A}$, in the range  
$g^{\rm eff}_{\rm A}=0.57-0.76$, were obtained. Around the same time
a similar study as in \cite{Faessler2008} was carried out in 
\cite{Suhonen2013c,Suhonen2014}
with results comparable with those of \cite{Faessler2008}. 

\begin{figure}[htb]
\includegraphics[scale=0.70]{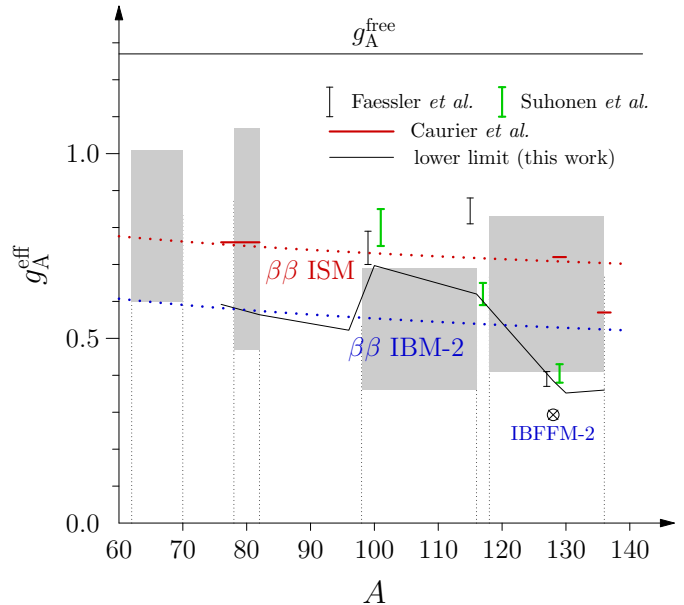}
\caption{Effective values of $g_{\rm A}$ in different theoretical approaches for
the nuclear mass range $A=62-136$. For more information see the text.}
\label{fig:gat}
\end{figure}

Later, the more
comprehensive pnQRPA studies \cite{Ejiri2015,Pirinen2015,Deppisch2016} of Gamow-Teller
$\beta$ decays of a wide range of nuclei ($A=62-142$) predicted heavily quenched values
of $g_{\rm A}$. These values depend on the mass number $A$ and they are depicted for
four separate mass ranges as hatched boxes in Fig.~\ref{fig:gat}. In the same figure,
with an obvious notation, also the curves (\ref{eq:curves}) are given, the IBM-2 (ISM)
results by a blue (red) dotted curve. Fig.~\ref{fig:gat} presents also the ISM results
of \cite{Caurier2012} (red horizontal bars), the pnQRPA results of \cite{Faessler2008}
(thin black vertical bars), and the pnQRPA results of 
\cite{Suhonen2013c,Suhonen2014} (thick green vertical bars). 
The cross inside a circle presents
the result $g^{\rm eff}_{\rm A}=0.293$ of a $\beta\beta$ analysis of $^{128}$Te using the 
microscopic interacting boson-fermion model (IBFFM-2) \cite{Yoshida2013}.

\section{Nuclear model and the two-stage fitting procedure \label{sec:model}}

There are many models which have recently been used to compute the $0\nu\beta\beta$ NMEs:
the quasiparticle random-phase approximation (QRPA), in its proton-neutron 
version (pnQRPA) (see \cite{Suhonen2012c} and references therein) and its 
renormalized extensions \cite{Toivanen1995,Raduta2010}, the interacting shell model 
(ISM) \cite{Caurier2005}, the microscopic interacting boson model (IBA-2)
\cite{Barea2009}, the Gogny-based energy-density functional (EDF)
\cite{Rodriguez2010} and the projected Hartree-Fock-Bogoliubov mean-field
scheme (PHFB) \cite{Rath2010}. Very recently
also the beyond-mean-field covariant density functional theory \cite{Yao2015} and
the corresponding non-relativistic version \cite{Vaquero2013} have been used to
describe the $0\nu\beta\beta$ decays of nuclei. For more details see the reviews
\cite{Engel2017,Suhonen2012d}.

The pnQRPA stands out favorably from the rest of the above-listed theoretical
approaches by the following features:
\begin{itemize}
\item[(i)] The pnQRPA is the only nuclear model which avoids the use of closure
approximation for the medium-heavy and heavy nuclei in the $0\nu\beta\beta$ 
calculations. However, there exists a non-closure calculation in the framework of
the ISM in \cite{Senkov2013} for the $0\nu\beta\beta$ decay of the light 
nucleus $^{48}$Ca.
\item[(ii)] pnQRPA can accommodate large single-particle bases, including all the
relevant spin-orbit-partner orbitals, in the calculations, 
\item[(iii)] pnQRPA may fail to predict properties of individual states but the
gross features of an energy region of nuclear states can be reliably accounted for
by the pnQRPA \cite{Civitarese1991}.
\end{itemize}
The features (i)--(iii) of the pnQRPA make it an ideal nuclear model to combine the
$2\nu\beta\beta$ and $0\nu\beta\beta$ calculations in a consistent way. Only for light
nuclei the single-particle model space is not dense enough for a reliable NME calculation,
in particular if the nucleus is doubly magic, like $^{48}$Ca, since the BCS
procedure used in pnQRPA framework is not very precise at closed major shells. This is
why the $0\nu\beta\beta$ decay of $^{48}$Ca is not included in the present analyses.
The above features (i)-(iii) were highlighted
in \cite{Rodin2006} where the $g_{\rm pp}$ parameter of the pnQRPA was
determined by reproducing the experimental $2\nu\beta\beta$ half-life for a given
value of $g_{\rm A}$, in that case $g_{\rm A}=g^{\rm free}_{\rm A}$. This approach was adopted
in the follow-up works of \cite{Kortelainen2007b,Kortelainen2007c,Simkovic2008} in
order to determine the Hamiltonian parameters as reliably as possible for a
consistent prediction of the $0\nu\beta\beta$ NMEs.

At this point it is fair to state that the ISM is a more complete theory than pnQRPA
in terms of the number of included many-body configurations in a given single-particle
model space. It has, at least, some approximate means to accommodate the feature (ii)
in the calculations, but the feature (i) is out of reach for the heavier nuclei. The ISM
has, however, the advantage that also it can produce $2\nu\beta\beta$ NMEs without
resorting to the closure approximation. This means that the analyses of the present
work could also be performed in the ISM framework when suitable ``particle-particle
interaction strength'', $g_{\rm pp}$(ISM), can be identified and fitted by the $2\nu\beta\beta$
half-life data through different values of $g_{\rm A}$. In fact, this type of study would
be highly welcome to see how robust the results of the present work really are.

The relation of the pnQRPA Hamiltonian and the $2\nu\beta\beta$ decay was further deepened
in the work \cite{Simkovic2013} where an isospin-restoration scheme for the pnQRPA 
was proposed. There the particle-particle parts of the pnQRPA matrices were divided into
isoscalar ($T=0$) and isovector ($T=1$) parts, splitting the particle-particle
interaction strength parameter $g_{\rm pp}$ to its isovector $g_{\rm pp}^{T=1}$ and 
isoscalar $g_{\rm pp}^{T=0}$ components. The two strength parameters can be adjusted separately
since they are practically independent of each other. The isovector parameter $g_{\rm pp}^{T=1}$
can be adjusted such that the Fermi NME of $2\nu\beta\beta$ decay vanishes leading to
the needed restoration of the isospin symmetry for the $2\nu\beta\beta$ decay. Keeping
the obtained value of $g_{\rm pp}^{T=1}$ one can independently vary $g_{\rm pp}^{T=0}$ in order
to fix its value by the data on $2\nu\beta\beta$-decay half-lives. The obtained
values of the two parameters can now, in turn, be used in the calculation of the 
$0\nu\beta\beta$ NMEs. Hence, in this two-stage fit the maximum amount of information
on $2\nu\beta\beta$ decay can be utilized to produce an optimized model Hamiltonian 
for the description of the $0\nu\beta\beta$ decay. This two-step fit was later used in
\cite{Hyvarinen2015} for pnQRPA-based $0\nu\beta\beta$-decay calculations and,
about the same time, in \cite{Barea2015} the isospin-restoration scheme was 
adopted for the IBM-2 $\beta\beta$ framework.

\section{Results for the reduced half-lives of the $0\nu\beta\beta$-decay
candidates \label{sec:results}}

\begin{figure}[htb]
\includegraphics[scale=0.75]{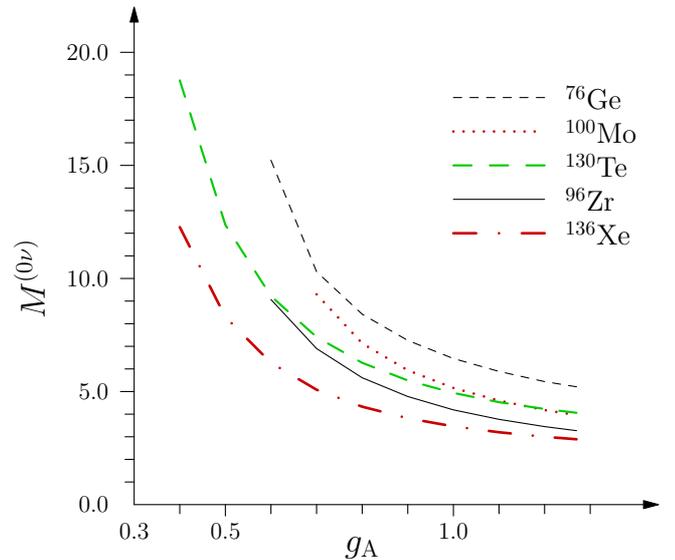}
\caption{Values of the $0\nu\beta\beta$ NME (\ref{eq:NME}) as functions 
of $g_{\rm A}$ for the decays of $^{76}$Ge, $^{96}$Zr, $^{100}$Mo, $^{130}$Te, and $^{136}$Xe.}
\label{fig:NMEs}
\end{figure}

In \cite{Simkovic2013,Hyvarinen2015} the two-step fit was quite limited: the experimental
$2\nu\beta\beta$ NME was extracted only for the free value $g^{\rm free}_{\rm A}=1.27$ and for a
moderately quenched value $g_{\rm A}=1.0$ for the axial-vector coupling constant. This
corresponds to two values of $g_{\rm pp}^{T=0}$ which were used in the 
$0\nu\beta\beta$-decay calculations. In this article the calculations of 
\cite{Hyvarinen2015} are extended to include all values of $g_{\rm A}$ below its
free value. The adopted single-particle bases, pairing parameters, and values of 
the isovector parameter $g_{\rm pp}^{T=1}$ are taken from \cite{Hyvarinen2015} (see, e.g.,
Table I of \cite{Hyvarinen2015}) and the details are not repeated here. Now, instead of
the two values of the $0\nu\beta\beta$ NMEs of \cite{Hyvarinen2015}, one obtains a
continuous sequence of $0\nu\beta\beta$ NMEs as a function of $g_{\rm A}$, as shown
for five exemplary cases in Fig.~\ref{fig:NMEs}. 

A general feature of the $0\nu\beta\beta$ NMEs, visible in Fig.~\ref{fig:NMEs},
is that they grow in magnitude with diminishing value of $g_{\rm A}$.
This stems from the structure (\ref{eq:NME}) of the $0\nu\beta\beta$ NME: The Fermi
and tensor NMEs are essentially independent of the value of $g_{\rm A}$, whereas the
Gamow-Teller NME grows in magnitude as $g_{\rm A}$ decreases and, at the same time, 
$g_{\rm pp}^{T=0}$ decreases (the magnitude of the computed $2\nu\beta\beta$ NME increases when
$g_{\rm pp}^{T=0}$ decreases, necessitating a decrease in the value of $g_{\rm A}$ in the
fit procedure). Thus the $g_{\rm A}$ dependence of the Gamow-Teller NME is not explicit
but induced by the $g_{\rm pp}$ dependence of the nuclear wave functions and the adopted
fit prescription of the $g_{\rm pp}^{T=0}$ parameter.
The Fermi NME adds coherently with the Gamow-Teller one since the NMEs
have opposite sign. In addition, the magnitude of the combination
$(g_{\rm V}/g_{\rm A})^2M^{(0\nu)}_{\rm F}$ grows large, comparable to the magnitude of
the Gamow-Teller NME, since the prefactor $(g_{\rm V}/g_{\rm A})^2$
grows rapidly with decreasing $g_{\rm A}$. 

The maximum value of the $2\nu\beta\beta$ Gamow-Teller NME
is attained at $g_{\rm pp}^{T=0}=0$. This, in turn, means that there is a minimum value of
$g_{\rm A}$ for which the maximum NME can fit the $2\nu\beta\beta$-decay half-life. This
minimum value of the possible $g_{\rm A}$ is an interesting by-product of the two-stage
fit and this lower limit of possible $g_{\rm A}$ values is presented in 
Fig.~\ref{fig:gat} as a solid black line and in Table~\ref{tab:ratios} in the
second column. From Fig.~\ref{fig:gat} it is seen that the here obtained lower limit of 
$g_{\rm A}$ is consistent with the thick green vertical bars of $g_{\rm A}$ ranges obtained in 
\cite{Suhonen2013c,Suhonen2014} and also consistent with the thin black vertical bars 
obtained in \cite{Faessler2008} for $A=100$ and $A=128$. For $A=116$ the analysis of
\cite{Faessler2008} produces a larger value of $g_{\rm A}$ than that obtained in
\cite{Suhonen2013c,Suhonen2014}. This difference could be due to different 
experimental $\log ft$ values for the EC branch $^{116}\textrm{In}\to\,^{116}\textrm{Cd}$
used in the different analyses, \cite{Suhonen2013c,Suhonen2014} using more
recent data for this decay branch than \cite{Faessler2008} (see \cite{Suhonen2014} 
for details).

There is a notable tension with the results of \cite{Faessler2008,Suhonen2013c,Suhonen2014}
and the the Gamow-Teller $\beta$-decay analysis 
(hatched region) in the interval $A=98-116$. The basic difference between the Gamow-Teller 
analyses of \cite{Ejiri2015,Pirinen2015,Deppisch2016} and the analyses of
\cite{Faessler2008,Suhonen2013c,Suhonen2014} is that the latter exploit the half-life
data on $2\nu\beta\beta$ decays and the former not. It is more appropriate to include
the $2\nu\beta\beta$ data if $0\nu\beta\beta$ properties are to be predicted since the
$2\nu\beta\beta$-decay observables are the closest to the $0\nu\beta\beta$-decay
observables. Thus, from the point of view of the present work it is preferable to compare
the results with those obtained in \cite{Faessler2008,Suhonen2013c,Suhonen2014}.

\begin{figure}[htb]
\includegraphics[scale=0.75]{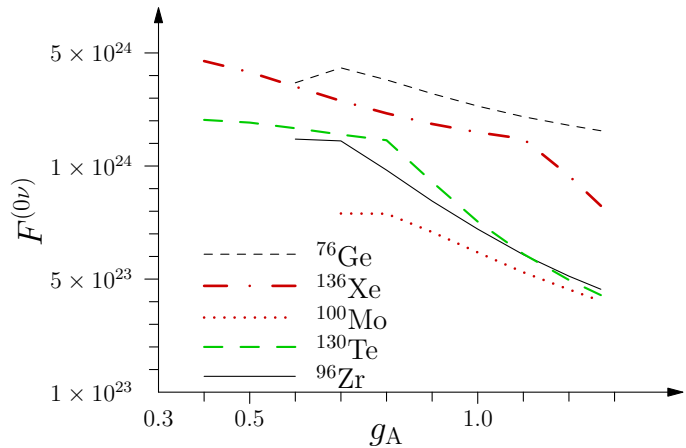}
\caption{Values of the reduced half-lives $F^{(0\nu)}$ as functions of $g_{\rm A}$ 
for the decays of $^{76}$Ge, $^{96}$Zr, $^{100}$Mo, $^{130}$Te, and $^{136}$Xe.}
\label{fig:Fs}
\end{figure}

As mentioned earlier, one would expect that the reduced half-life $F^{(0\nu)}$ of 
(\ref{eq:F}) would grow enormously with increasing value of $g_{\rm A}$ since 
it is proportional to $(g_{\rm A})^{-4}$. This is in the very core of the large impact 
caused by the work by Barea \textit{et al.} in \cite{Barea2013}. Since 
now all the ingredients of (\ref{eq:F}) in the two-stage fit have been produced, 
it is possible 
to plot the values of $F^{(0\nu)}$ as functions of $g_{\rm A}$. This has been done in
Fig.~\ref{fig:Fs} for the exemplary cases whose NMEs were presented in 
Fig.~\ref{fig:NMEs}. Inspection of Fig.~\ref{fig:Fs} confirms a rather unexpected
stunning result which is the reason for this article: The reduced half-lives are
not much affected by the large variation in the values of $g_{\rm A}$. The maximum
and minimum values of this half-life, as also their ratios, are listed in 
Table~\ref{tab:ratios}. From the table one can see that the ratio of the
reduced (and also actual) half-lives ranges from 2.0 (the decay of $^{100}$Mo) to
5.7 (the decay of $^{136}$Xe). This is extremely good news for the present and
future $0\nu\beta\beta$-decay experiments since even the smallest values of $g_{\rm A}$,
obtained in some nuclear models, do not ruin the sensitivity of the experiments.

\begin{table}
\caption{\label{tab:ratios} The obtained minimum values of $g_{\rm A}$, the computed 
maximum and minimum values of the reduced
half-life $F^{(0\nu )}$, and their ratio for the studied $0\nu\beta\beta$ systems.}
\begin{ruledtabular}
	\begin{tabular}{lcccc}
& \multicolumn{2}{c}{$F^{(0\nu )}$} & \\
System & $g_{\rm A}$(min) & $F^{(0\nu )}$(max) & $F^{(0\nu )}$(min) & Ratio \\
	\hline
$A=76$ & 0.59 & $4.4\times 10^{24}$ & $1.6\times 10^{24}$ & 2.7 \\
$A=82$	& 0.56 & $2.1\times 10^{24}$ & $7.0\times 10^{23}$ & 3.0 \\
$A=96$	& 0.52 & $2.2\times 10^{24}$ & $4.5\times 10^{23}$ & 4.9 \\
$A=100$	& 0.70 & $7.9\times 10^{23}$ & $4.0\times 10^{23}$ & 2.0 \\
$A=116$	& 0.62 & $1.1\times 10^{24}$ & $3.6\times 10^{23}$ & 3.1 \\
$A=128$	& 0.38 & $3.4\times 10^{25}$ & $7.7\times 10^{24}$ & 4.4 \\
$A=130$	& 0.35 & $2.1\times 10^{24}$ & $4.3\times 10^{23}$ & 4.9 \\
$A=136$	& 0.36 & $4.7\times 10^{24}$ & $8.2\times 10^{23}$ & 5.7 \\
	\end{tabular}
\end{ruledtabular}
\end{table}

\section{Conclusions \label{sec:conclusions}}

In conclusion, in this article it is shown by using a two-stage fit procedure to
two-neutrino $\beta\beta$-decay data in the pnQRPA nuclear-theory framework that even
the very strongly quenched values of the axial-vector coupling constant, obtained
in many theoretical analyses, decrease the sensitivity of the present and future
neutrinoless $\beta\beta$-decay experiments only by reduction factors $R_1=2-6$, depending
on the decaying nucleus. This is in strong contrast with the possible huge values of the 
usually assumed simple-minded reduction factor $R_2=(g^{\rm free}_{\rm A}/g^{\rm eff}_{\rm A})^4$, 
containing the ratio of the free and effective values of the axial-vector coupling constant:  
values of $R_2\approx 300$ are reachable in the worst scenarios proposed. 
This means that the knowledge of the effective value of
the axial-vector coupling constant is not crucial for the neutrinoless $\beta\beta$-decay
experiments from the point of view of experimental sensitivity.

The reason behind the moderate reduction in the experimental sensitivity is the strong 
dependence of the $\beta\beta$-decay NME on the value of the axial-vector coupling constant 
through the two-step fit procedure exploiting the two-neutrino $\beta\beta$-decay data. This
mechanism is behind the important new finding that is communicated here, offering
much better perspectives for detecting the neutrinoless $\beta\beta$ decay.
Thus, the presently obtained results are extremely important for the community of 
$\beta\beta$ experiments, as also for nuclear, neutrino, and particle physics in general.

\section*{Acknowledgments}

This work has been partially supported by the 
Academy of Finland under the Finnish Centre of Excellence Programme 
2012-2017 (Nuclear and Accelerator Based Programme at JYFL).

\end{document}